\documentclass[fleqn,usenatbib,useAMS]{mnras}
\usepackage{savesym}
\savesymbol{tablenum}
\usepackage{siunitx}
\restoresymbol{SIX}{tablenum}
\usepackage{graphicx,url,amssymb,amsmath,color,units,wasysym,epsfig,epstopdf,xcolor,soul,xcolor,subfigure}
\usepackage{verbatim}
\usepackage{empheq}
\usepackage[normalem]{ulem}

\usepackage{newtxtext,newtxmath}
\newcommand{\new}[1]{#1}

\newcommand{\seobnre}{{\texttt{SEOBNRE} }}
\pubyear{2020}

\begin{document}
\label{firstpage}
\pagerange{\pageref{firstpage}--\pageref{lastpage}}

\title[On the origin of GW190425]{On the origin of GW190425}

\author[Romero-Shaw et al.]{Isobel M. Romero-Shaw$^{1,2}$, Nicholas Farrow$^{1,2}$, Simon Stevenson$^{3,2}$, \newauthor Eric Thrane$^{1,2}$, Xing-Jiang Zhu$^{1,2}$\thanks{E-mail: zhuxingjiang@gmail.com}\\
$^{1}$School of Physics and Astronomy, Monash University, Clayton, VIC 3800, Australia\\
$^{2}$OzGrav: The ARC Centre of Excellence for Gravitational Wave Discovery, Australia\\
$^{3}$Centre for Astrophysics and Supercomputing, Swinburne University of Technology, Hawthorn, VIC 3122, Australia
}

\maketitle

\begin{abstract}
The LIGO/Virgo collaborations recently announced the detection of a binary neutron star merger, GW190425.
The mass of GW190425 is significantly larger than the masses of Galactic double neutron stars known through radio astronomy.
We hypothesize that GW190425 formed differently than Galactic double neutron stars, via unstable ``case BB'' mass transfer.
According to this hypothesis, the progenitor of GW190425 was a binary consisting of a neutron star and a ${\sim}4$--$\unit[5]{M_\odot}$ helium star, which underwent common-envelope evolution.
Following the supernova of the helium star, an eccentric double neutron star was formed, which merged in ${\lesssim}\unit[10]{Myr}$.
The helium star progenitor may explain the unusually large mass of GW190425, while the short time to merger may explain why similar systems are not observed in radio.
To test this hypothesis, we measure the eccentricity of GW190425 using publicly available LIGO/Virgo data.
We constrain the eccentricity at $\unit[10]{Hz}$  to be $e \leq 0.007$ with $90\%$ confidence.
This provides no evidence for or against the unstable mass transfer scenario, because the binary is likely to have circularized to $e\lesssim10^{-4}$ by the time it was detected.
Future detectors will help to reveal the formation channel of mergers similar to GW190425 using eccentricity measurements.
\end{abstract}


\section{Introduction}
Gravitational waves produced by a binary neutron star (BNS) merger have been detected for the second time~\citep{gw170817, GWTC1, gw190425} by Advanced LIGO~\citep{AdvancedLIGO} and Virgo~\citep{AdvancedVirgo}. 
The binary GW190425 is remarkable because it is significantly more massive than Galactic BNS~\citep{gw190425}.
%\irs{Total mass 2.875 Msun pm 0.0014 vs 3.3 Msun pm 0.1}
%\irs{Chirp mass 1.24 Msun (errors?) vs 1.4866 pm 0.0002)
Of the 17 Galactic BNS with reported mass measurements \citep[see][and references therein]{FarrowDNSmass}, the most massive has total mass $M = 2.886 \pm 0.001 M_{\odot}$~\citep{Lazarus16,Ferdman17}. 
For GW190425, $M = 3.4^{+0.3}_{-0.1} M_{\odot}$, which is inconsistent with the observed Galactic population \citep{gw190425}. 
This invites speculation about its formation channel. 

BNS may form through isolated binary evolution~\citep{SmarrBlandford, Srinivasan89, PortgiesZwart, Canal, Kalogera06, Postnov14tza, Beniamini15, Vigna-Gomez18d, Kruckow18, Giacobbo18-1, Giacobbo18-2, Giacobbo19, Mapelli18} or through dynamical interactions~\citep{PhinneySigurdsson91, Sigurdsson94ju, Kuranov2006, Ivanova07pm, Kiel10, Benacquista11kv, East12ww, Palmese17, Andrews19}.
The dominant formation channel for Galactic BNS is thought to be isolated evolution: a stellar binary in the field experiences successive supernovae, and the stellar remnant of each component is a neutron star~\citep{Tauris17, Vigna-Gomez18d}.
%The discrepancies between the mass of GW190425 and the observed Galactic BNS suggest two different formation channels.
While many neutron stars {\em not} in BNS are known to have masses consistent with the components of GW190425 \citep{Ozel16,Alsing18}, the high mass of this system in not easily explained by standard isolated evolution, since the large supernova kicks associated with massive NS formation are expected to disrupt binaries; see \citet{Michaely16}, and references therein.

In the dynamical formation case, BNS form through interactions inside dense stellar environments, such as globular clusters.
A NS, which may have a stellar companion, sinks to the cluster core through dynamical friction.
This can only occur once the number of black holes in the core has been depleted, either due to merger-induced kicks or because they gain velocity through dynamical interactions \citep[e.g.,][]{Breen13}. 
In the core, the NS preferentially swaps any existing stellar companion for another NS, forming a BNS with a short merger time \citep{Zevin19ns}.
% However, the merger rate of dynamically-formed BNS is expected to be low relative to the total local BNS merger rate.
\new{While the dynamical hypothesis provides an explanation for the large mass of GW190425, it is difficult to reconcile the implied merger rate with that predicted by $N$-body simulations~\citep{Grindlay05, Bae13, Belczynski17mqx, Ye20DynamicBNSrate}; see also \citet{Papenfort}, and references therein. 
Current estimates sit at around $0.003$---$\unit[6]{Gpc^{-3} yr^{-1}}$~\citep{Tsang}, which, for advanced LIGO's BNS range of $\sim\unit[100]{Mpc}$~\citep{AdvancedLIGO}, translates to a predicted rate of $1.25\times10^{-5}$---$\unit[2.5\times10^{-2}]{yr^{-1}}$.
For a different perspective see \citet{Andrews19}, who highlight that tight and highly-eccentric Galactic-field BNS may form dynamically, provided that their host clusters have sufficiently high central densities.}

\begin{figure}
    \centering
    \includegraphics[scale=0.37]{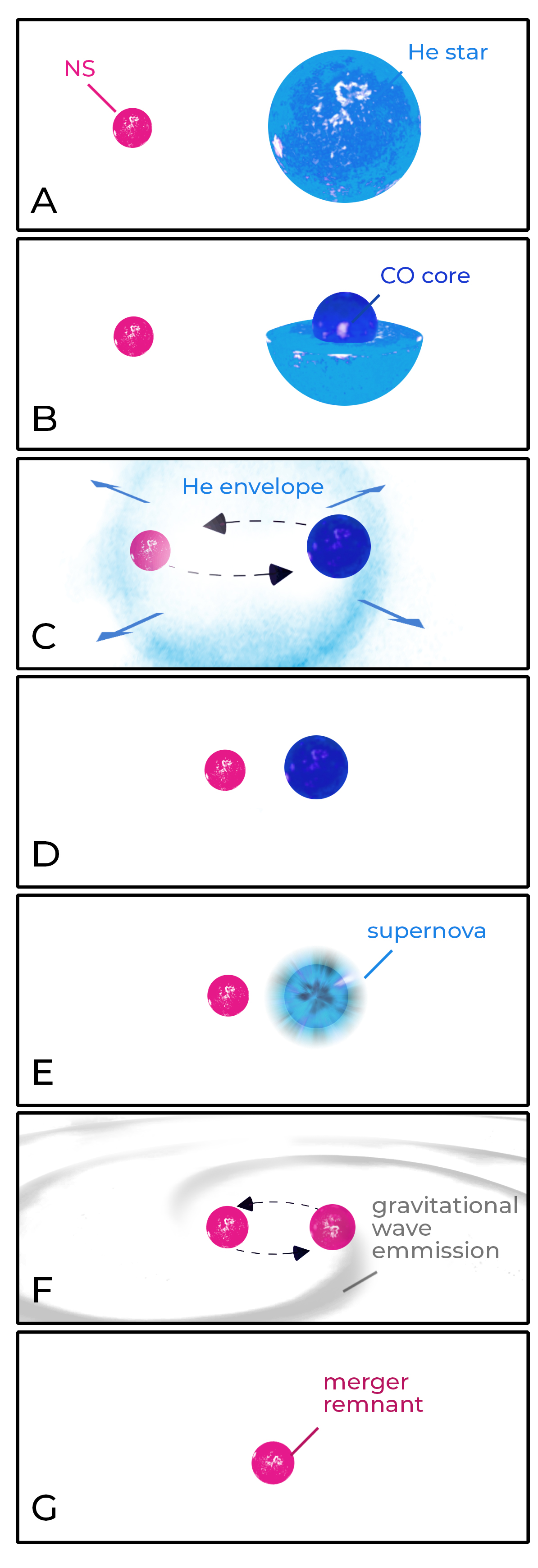}
    \caption{Illustration of unstable case BB mass transfer leading to a BNS merger. Credit: Carl Knox.}
    \label{fig:unstable_case_bb_mt}
\end{figure}

We argue that massive BNS like GW190425 may evolve in isolation if they undergo a process known as unstable ``case BB'' mass transfer (MT)~\citep{Delgado81, Tutukov93b, Tutukov93a, Zwart96, PortgiesZwart, Belczynski02, Belczynsk01uc, Dewi02, Ivanova03, Zevin19ns}. 
We illustrate this process in Fig.~\ref{fig:unstable_case_bb_mt}. 
The He star companion of a NS (panel A) fills its Roche lobe after the end of its He core burning phase (panel 
B), initiating common-envelope evolution (panel C).
The He envelope is ejected, leaving behind a NS--CO core binary (panel D) that is tight enough to survive the supernova of the He star (panel E).
The resulting BNS inspirals due to emission of gravitational waves (panel F) and eventually merges, leaving behind a NS or black hole remnant (panel G).
Unstable case BB MT may produce heavy BNS with unequal masses~\citep{Ivanova03, Mueller16}.
The supernova kick can also leave the binary with significant eccentricity \citep{Brandt1995}, which can act as an identifier for this formation channel.
During standard isolated evolution, gravitational radiation gradually circularizes binaries before they get close to merger \citep{Peters1964, Hinder08}.
% However, unstable case BB MT may produce an eccentric binary that merges before it has time to completely circularize.
On the other hand, as we show in Sec.~\ref{sec:formation} of this work, binaries formed through unstable case BB MT have eccentricities $10^{-6} \lesssim e \lesssim 10^{-3}$ when they enter the LIGO/Virgo band.
%, with the peak of the distribution at $e \sim 10^{-4.5}$.

GW190425 was detected by a search algorithm that assumes quasi-circular binary orbits.  
Its properties, presented in \cite{gw190425}, were inferred by matched-filtering data against quasi-circular waveform models.
Burst searches may flag eccentric signals, but cannot measure their eccentricity~\citep[e.g.,][]{Salemi19}.  
Recently, \citet{Nitz19} performed a matched-filtering search for eccentric BNS signals using inspiral-only eccentric waveform models. 
Computationally efficient, inspiral-merger-ringdown models of eccentric waveforms are not yet available, although development is ongoing \citep[e.g.,][]{Huerta17, Tiwari19}.
Computationally inefficient models \citep[e.g.,][]{SEOBNRE} take too long to generate to be used for straightforward Bayesian inference, which relies on $\mathcal{O}(100)$ waveform computations per iteration of its sampling algorithm.
We can, however, use such models to efficiently obtain eccentricity measurements by post-processing the posterior probabilities for quasi-circular waveform models, as demonstrated in \citet{Romero-Shaw19}; see also~\cite{Lower}.

In this Letter, we take steps towards identifying the formation channel of GW190425 using orbital eccentricity measurements.
We simulate BNS evolving through unstable case BB MT and compare the resulting eccentricity distribution to the posterior probability on eccentricity for GW190425.
The remainder of this work is structured as follows.
In Sec.~\ref{sec:formation}, we describe unstable case BB MT and outline our method for simulating the expected eccentricity distribution at $\unit[10]{Hz}$ from this channel.
We present upper limits on the orbital eccentricity of GW190425 when its gravitational radiation has a frequency of $\unit[10]{Hz}$ in Sec.~\ref{sec:eccentricity}, comparing the posterior probability distribution to the eccentricity distribution expected from supernova kicks.
We discuss the implications of our results for the formation pathway of GW190425 in Sec.~\ref{sec:discussion}.

\section{The isolated evolution of GW190425}
\label{sec:formation}
\subsection{Unstable mass transfer in the isolated binary evolution channel}
\label{sec:formation:a}

The immediate progenitor of an isolated BNS is a binary comprising a NS and a helium (He) star with orbital period $\unit[{\sim} 0.1 - 2]{days}$, which has evolved thus far via common-envelope (CE) evolution~\citep{Belczynsk01uc, Dewi02, Ivanova03, Ivanova13, Zevin19ns}. 
The He star then expands, filling its Roche lobe, and transferring mass onto the NS.
If the mass transfer process is unstable, it can lead to a second CE (2CE) phase \citep{Ivanova03,DewiPols03}. 
The surviving post-2CE system consists of the carbon-oxygen (CO) core of the He star and the original NS. 
The latter has accreted only a small amount of mass ($\sim 0.05 - 0.1 M_{\odot})$~\citep{MacLeod14yda} during the 2CE phase. 
The binary can be tight enough that its orbital period is $< \unit[1]{hr}$, making it likely to survive the subsequent supernova explosion of the CO core. 

This asymmetric supernova explosion gives the compact object a kick.
In population synthesis studies, kick velocities are often assumed to follow Maxwellian distributions.
Core collapse supernovae are thought to produce large kicks, with one-dimensional standard deviation $\sigma\approx\unit[265]{km\,s^{-1}}$ \citep{Hobbs2005}, while ultra-stripped supernovae and electron-capture supernovae are thought to produce small kicks, $\sigma\approx\unit[30]{km\, s^{-1}}$ \citep{Vigna-Gomez18d, Giacobbo18-2, Giacobbo19}.

The relationship between the final He star mass (CO core mass) and the NS remnant mass is uncertain, but \citet{Mueller16} predict that a ${\sim}\unit[4-5]{M_{\odot}}$ He star (with a ${\sim} 3 M_{\odot}$ CO core) corresponds to a ${\sim} 2 M_{\odot}$ NS \citep[see also][]{Tauris2015}. It is assumed that there is an instantaneous mass loss of ${\sim} 1 M_{\odot}$ during supernova.
If the pre-2CE binary consists of a ${\sim} 1.4 M_{\odot}$ NS and a ${\sim} 4 - 5 M_{\odot}$ He star, then the post-2CE, post-supernova binary is a ${\sim} (1.4 + 2.0) M_{\odot}$ BNS which merges in $<\unit[10]{Myr}$.
BNS with this lifespan are far less likely to be detected in radio pulsar surveys than their longer-lived counterparts, and BNS with orbital periods $< \unit[1]{hr}$ are effectively invisible in current pulsar searches. 
For example, the acceleration search of \citet{Cameron17BNS}, which found the most accelerated pulsar observed to date, was sensitive to binary pulsars with orbital periods down to $\unit[1.5]{hr}$.
We discuss selection effects further in Appendix \ref{sec:selectioneffect}.

\subsection{Eccentricity distribution}
\label{sec:formation:b}
Following Equations 2.1 to 2.8 from ~\citet{Brandt1995}, we calculate eccentricities introduced by supernova kicks in this formation scenario. 
% We define the dimensionless mass $\tilde{m}$ as
% \begin{align}
%     %\tilde{m}=\frac{m_1+m_2}{m_1'+m_2} && \tilde{v}=\frac{v_{\textrm{kick}}}{v_{\textrm{orbit}}}
%     \tilde{m}=\frac{M_{1}^{(\textrm{CO})}+M_{2}^{(\textrm{NS})}}{M_{1'}^{(\textrm{NS})}+M_{2}^{(\textrm{NS})}},
% \end{align}
% where $M_{1}^{(\textrm{CO})}$ is the mass of the progenitor CO core, $M_{1'}^{(\textrm{NS})}$ is the mass of the neutron star remnant, and $M_{2}^{(\textrm{NS})}$ is the mass of the companion NS. 
% We define the dimensionless kick velocity $\tilde{v}$ as
% \begin{align}
%   \tilde{v}=\frac{v_{\textrm{kick}}}{v_{\textrm{orbit}}}.
% \end{align}
% The resultant orbital eccentricity is
% \begin{equation}
%     \label{eccentricityEqn}
%     \begin{aligned} 
%         e = & \left[1-\tilde{m} \left(2-\tilde{m}\left(1+2 \tilde{v} \cos \phi \cos \theta+\tilde{v}^{2}\right)\right) \right. \\ 
%             & \left. \times\left((1+\tilde{v} \cos \phi \cos \theta)^{2}+(\tilde{v} \sin \theta)^{2}\right) \right]^{\frac{1}{2}},
%     \end{aligned}
% \end{equation}
% where $\phi$ is the angle between the direction of the star's velocity vector and the kick vector, and $\theta$ is the angle between the orbital plane and the kick.
% The orbit acquires a new semimajor axis $a'$ from the supernova kick, which is given by
% \begin{equation}
%     \label{newsemimajorEqn}
%     a' = \left[\frac{1}{a} \left(2 - \tilde{m} \left(1+2\tilde{v}\cos{\phi}\cos{\theta}+\tilde{v}^2\right)\right)\right]^{-1}.
% \end{equation}
We simulate binaries with first-born NS of mass $1.4\, M_{\odot}$ and CO core of mass $3.0\, M_{\odot}$, which lead to a second-born NS of mass $2.0\, M_{\odot}$, and draw orbital periods at time of supernova from a log-uniform distribution between $\unit[0.1]{hr}$ and $\unit[1]{hr}$~\citep[see Fig.~8 from][]{Vigna-Gomez18d}.
Supernova kick velocities are drawn from Maxwellian velocity distributions, with $\sigma=\unit[265]{km\,s^{-1}}$ for large kicks and $\sigma=\unit[30]{km\,s^{-1}}$ for small kicks.
We simulate isotropically-distributed kicks, and discard NS that receive kicks sufficient to disrupt the binary.
Each binary's eccentricity is evolved according to \citet{Peters1964} until its gravitational-wave frequency reaches $f_{\textrm{gw}}=\unit[10]{Hz}$.

\begin{figure}
\begin{center}
  \includegraphics[width=0.48\textwidth]{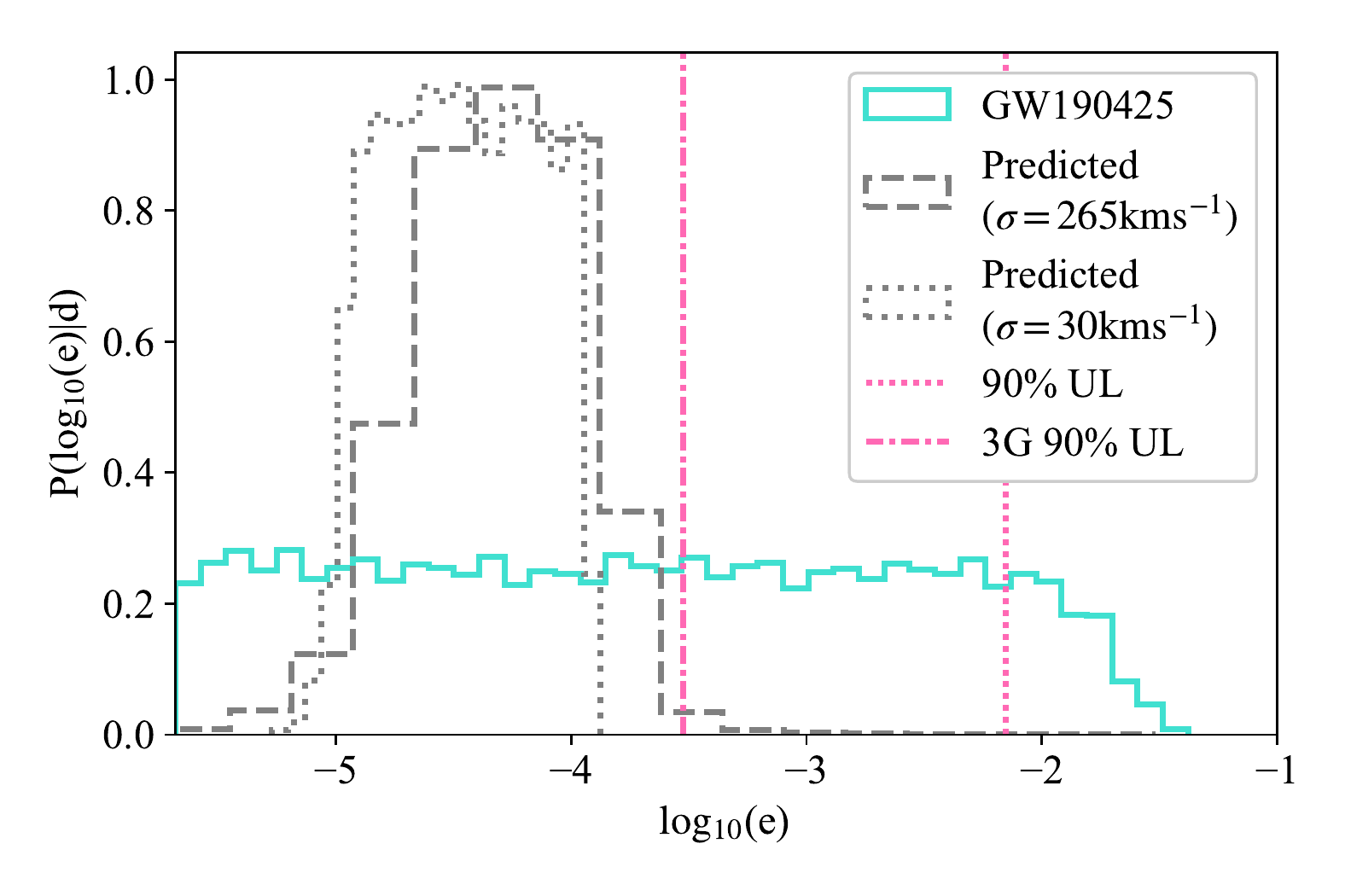}
  \caption{\new{Posterior distribution on $\log_{10}(e)$ for GW190425, alongside eccentricities acquired during unstable case BB MT from kicks with velocities drawn from Maxwellian distributions. 
  We indicate our measured 90\% confidence upper limit on the eccentricity of GW190425 at \unit[10]{Hz} with a dashed bar at $e = 0.007$, and our estimate of the third-generation detector network upper limit with a dot-dashed bar at $e=0.0003$.
  Space-based detector LISA will be able to resolve BNS eccentricities within the entire unstable case BB MT range; see Sec.~\ref{sec:discussion}, with reference to \citet{Lau19}.
  Our simulated eccentricity distributions agree with the subpopulation of ultra-compact BNS studied by \citet{Kowalska11}.}}
  \label{fig:combinedPlot}
\end{center}
\end{figure}

We present the distributions of $\log_{10}$ eccentricities obtained in this scenario in Fig.~\ref{fig:combinedPlot}.
Higher-velocity kicks tend to cause slightly higher eccentricities.
Measured at \unit[10]{Hz}, supernovae with large kicks lead to a $\log_{10}(e)$ distribution with a mean of $-4.30$, while the $\log_{10}(e)$ distribution arising from smaller kicks has a mean of $-4.46$.
The $90\%$ confidence interval on $\log_{10}(e)$ spans $-4.94 \leq \log_{10}(e) \leq -3.98$ for small kick velocities, and $-4.89 \leq \log_{10}(e) \leq -3.79$ for large kick velocities.

\section{Eccentricity of GW190425}
\label{sec:eccentricity}
To compare GW190425 to the model described in Sec.~\ref{sec:formation}, we measure its eccentricity when it enters the frequency band of LIGO/Virgo.
\citet{Romero-Shaw19} demonstrated the calculation of Bayesian posterior probability distributions for the eccentricity of binaries detected in gravitational waves; see~\cite{payne} for detailed formulation of the reweighting procedure that underlies such post-processing techniques.
Following the same method, we use circular waveform model \texttt{IMRPhenomD}~\citep{Khan15} to compute ``proposal'' posterior probability distributions for the binary parameters, and eccentric waveform model \seobnre~\citep{SEOBNRE} to reweight to our ``target'' distribution. 

We use the Bayesian inference library \texttt{BILBY}~\citep{ashton19} to fit the data to the proposal model.
Our prior on chirp mass $\mathcal{M}$ is uniform between $1.42$ and $2.60$ M$_{\odot}$, and our prior on mass ratio $q$ is uniform between $0.125$ and $1$. 
Our prior on source luminosity distance $d_\text{L}$ is uniform in co-moving volume between $1$ and $\unit[500]{Mpc}$. 
For dimensionless aligned component spins $\chi_1$ and $\chi_2$, we use priors that are uniform between $-1$ and $0.6$ due to limitations of the \seobnre waveform model. 
For the remaining sampled parameters -- right ascension, declination, source inclination angle, polarisation angle and reference phase -- we use standard priors.

To reduce the time spent evaluating computationally-expensive \seobnre waveforms, we make initial measurements at a reference frequency of \unit[20]{Hz}.
To obtain orbital eccentricity measurements at \unit[10]{Hz}, we follow  \citet{Peters1964} to evolve the system backwards in time.
We use a log-uniform prior on eccentricity, with a resolution of 30 bins per sample.
At \unit[20]{Hz}, our prior is in the range $-6 \leq \log_{10}(e) \leq -1$, translating to $-5.68 \leq \log_{10}(e) \leq -0.71$ at \unit[10]{Hz}.

We constrain the eccentricity of GW190425 to be $e~\leq~0.007$ at $\unit[10]{Hz}$ with 90\% confidence.
Our reweighting efficiency is $0.386$, giving us $7718$ effective samples; see~\cite{Romero-Shaw19} for a discussion of efficiency.
See Appendix \ref{sec:intrinsic} for posteriors on the intrinsic source parameters, which are consistent with results from~\cite{gw190425}.

We compare the eccentricity posterior for GW190425 to the eccentricity distribution expected from supernova kicks during unstable case BB MT in Fig.~\ref{fig:combinedPlot}.
\new{The posterior probability distribution for the eccentricity of GW190425 is consistent with our log-uniform prior for eccentricities $e \lesssim 7 \times 10^{-3}$ at $\unit[10]{Hz}$, implying that we are unable to resolve differences between eccentricities lower than this with existing instruments. 
While the eccentricity of GW190425 is consistent with eccentricities induced during unstable case BB MT, we cannot distinguish this channel from other mechanisms using eccentricity measurements obtained with advanced LIGO/Virgo.}

\section{Discussion}
\label{sec:discussion}
\new{We constrain the eccentricity of GW190425 to $e \leq 0.007$ at $\unit[10]{Hz}$.
GW190425 may have formed through unstable case BB MT, but with present-day detectors, we are unable to distinguish the small residual eccentricity expected from this channel at $\unit[10]{Hz}$.
Proposed third-generation observatories such as Cosmic Explorer~\citep[CE;][]{CosmicExplorer} and the Einstein Telescope~\citep[ET;][]{ET2010} will detect GW190425-like binaries with higher signal-to-noise ratios, and at lower frequencies.
Following the calculation outlined in \citet{Lower}, we find that a network of $2 \times$ CE can measure the eccentricity of a GW190425-like BNS if $e \geq 0.0003$ at $\unit[10]{Hz}$, and will be able to observe the upper tail of the distribution expected from high-velocity kicks.
The space-based gravitational-wave detector LISA~\citep{LISA} will be sensitive down to \unit[$10^{-4}$]{Hz}, enabling sub-categories of both isolated and dynamical mergers to be distinguished \citep{Breivik16, Nishizawa17, DOrazio18, Samsing18, Lau19}. 
From \citet{Lau19}, the resolvable eccentricity of LISA at $\sim \unit[1]{mHz}$ is $e \gtrsim 0.001$, which translates to $e \gtrsim 10^{-7}$ at $\unit[10]{Hz}$. 
Hence, the predicted eccentricity distribution from unstable case BB MT will be resolvable with LISA.}

The inferred merger rate for GW190425-like systems is high compared to lighter BNS \citep{gw190425}. 
Since GW190425 is only the second BNS merger to be observed in gravitational waves, roughly half of all BNS mergers may form by the same means. 
This could imply that unstable case BB MT is a common pathway to BNS formation. 
Any proposed formation channel for this merger must also explain the relatively high formation rate of similar BNS. 

Measurements of NS spins, which are imprinted on gravitational-wave signals through the effective spin parameter $\chi_\text{eff}$, can provide additional clues to the formation channel of BNS.
The $\chi_\text{eff}$ of Galactic-field BNS, thought to have formed via standard isolated evolution, are predicted to range from $0.00$ to $0.02$ at merger~\citep{Zhu17}.
\new{Although we have not devoted much discussion to the dynamical formation hypothesis for GW190425 because it is theoretically disfavoured, it remains possible that GW190425-like BNS can form dynamically in, for example, globular clusters.}
Such BNS can have a wider range of spins than their isolated counterparts \citep[see][and references therein]{East15}.
A binary with measurably negative $\chi_\text{eff}$ would be difficult to explain through anything other than dynamical formation.
The $\chi_\text{eff}$ of BNS formed through unstable case BB MT depends critically on the amount of angular momentum transferred onto the first-born NS during two CE stages, since the second-born NS is expected to spin down to effectively zero spin in a timescale comparable to the merger time ($\lesssim \unit[10]{Myr}$).
\citet{SoumiDe19} suggest that black holes tend to preserve their natal masses and spins during CE evolution. If this holds up for BNS, it might imply that all BNS formed through unstable case BB MT are expected to have low dimensionless component spins of $\chi<0.05$ at merger.
While we are unable to measure NS spins in GW190425 (and in GW170817), it may be possible to do so for future discoveries, allowing stronger constraints to be placed on system origins.

We note that, in systems with misaligned component spins, the signature of spin-induced precession in the signal can mimic the signature of eccentricity, leading to non-negligible eccentricity measurements for precessing quasi-circular binaries~(Romero-Shaw, Lasky \& Thrane, in preparation). We see no evidence of significant eccentricity in the signal of GW190425, so any degeneracy between precession and eccentricity does not influence our conclusion. Regardless, the in-plane spin of BNS is believed to be small \citep[e.g.,][]{FerdmanDPSR}, so this degeneracy is more important for binary black hole signals.

\section{acknowledgments}
\new{We thank Mike Lau for his insight into the eccentricity sensitivity of LISA, and Marcus Lower for letting us use his code to estimate the BNS eccentricity sensitivity of third-generation detectors.} This work is supported by Australian Research Council grants CE170100004 and FT150100281.
%This is LIGO document P1900321.
This research has made use of data, software and/or web tools obtained from the Gravitational Wave Open Science Center (https://www.gw-openscience.org), a service of LIGO Laboratory, the LIGO Scientific Collaboration and the Virgo Collaboration.

\textit{Note added}.-- While preparing this manuscript, we became aware of a pre-print claiming that the merger rate implied by GW190425 is inconsistent with population synthesis results for fast-merging BNS \citep{Note}. Similar discrepancies have arisen for GW170817, but studies show that predicted merger rates are consistent with observations when various model uncertainties, e.g., NS natal kicks, CE evolution, metallicity-specific star formation rate \citep{Chruslinska17, Giacobbo18-1, Belczynski17mqx, Chruslinska18, Neijssel19, Tang20}, are included. We therefore believe that we cannot rule out the formation of GW190425 through unstable case BB MT based solely on the inferred BNS merger rate.

\bibliographystyle{mnras}
\bibliography{ref}

\begin{appendix}
\section{Selection effects}
\label{sec:selectioneffect}
There are several selection effects that could cause discrepancy between the mass distribution of Galactic BNS observed in radio and that of extra-galactic BNS mergers detected in gravitational waves.
First, more massive binary mergers are detectable at further distances with gravitational waves. Assuming a uniform-in-comoving-volume source distribution, the \textit{observed} chirp mass distribution differs from the \textit{true} distribution by a factor of $\mathcal{M}^{5/2}$.
Second, more massive BNS merge faster, making them less likely to be discovered in pulsar surveys. However, the binary lifetime scales more strongly with its initial orbital period and eccentricity. As long as the binary chirp mass does not correlate \textit{strongly} with initial orbital period or eccentricity\footnote{Noting the mild correlation between the mass of second-born NS and orbital eccentricity for Galactic BNS \citep[e.g., Fig. 17 of][]{Tauris17}.}, the mass distribution of BNS observed in radio is a good representation of the birth distribution.
Third, the binary total masses ($M$) of Galactic BNS are known from measurements of the advance of periastron, which is proportional to $M^{2/3}$. This leads to a slight preference within the observed Galactic BNS sample towards higher total masses as well as shorter orbital periods, which make periastron advance and orbital decay rates easier to measure.
The fact that GW190425 is significantly more massive than all 17 knwon Galactic BNS may suggest an invisible Milky Way BNS population that is formed in ultra-tight, possibly highly eccentric orbits, as produced via unstable case BB MT.

 \section{Recovered posterior probability distributions for GW190425}
 \label{sec:intrinsic}

\begin{figure}
    \label{fig:intrinsic}
    \centering
    \includegraphics[scale=0.33]{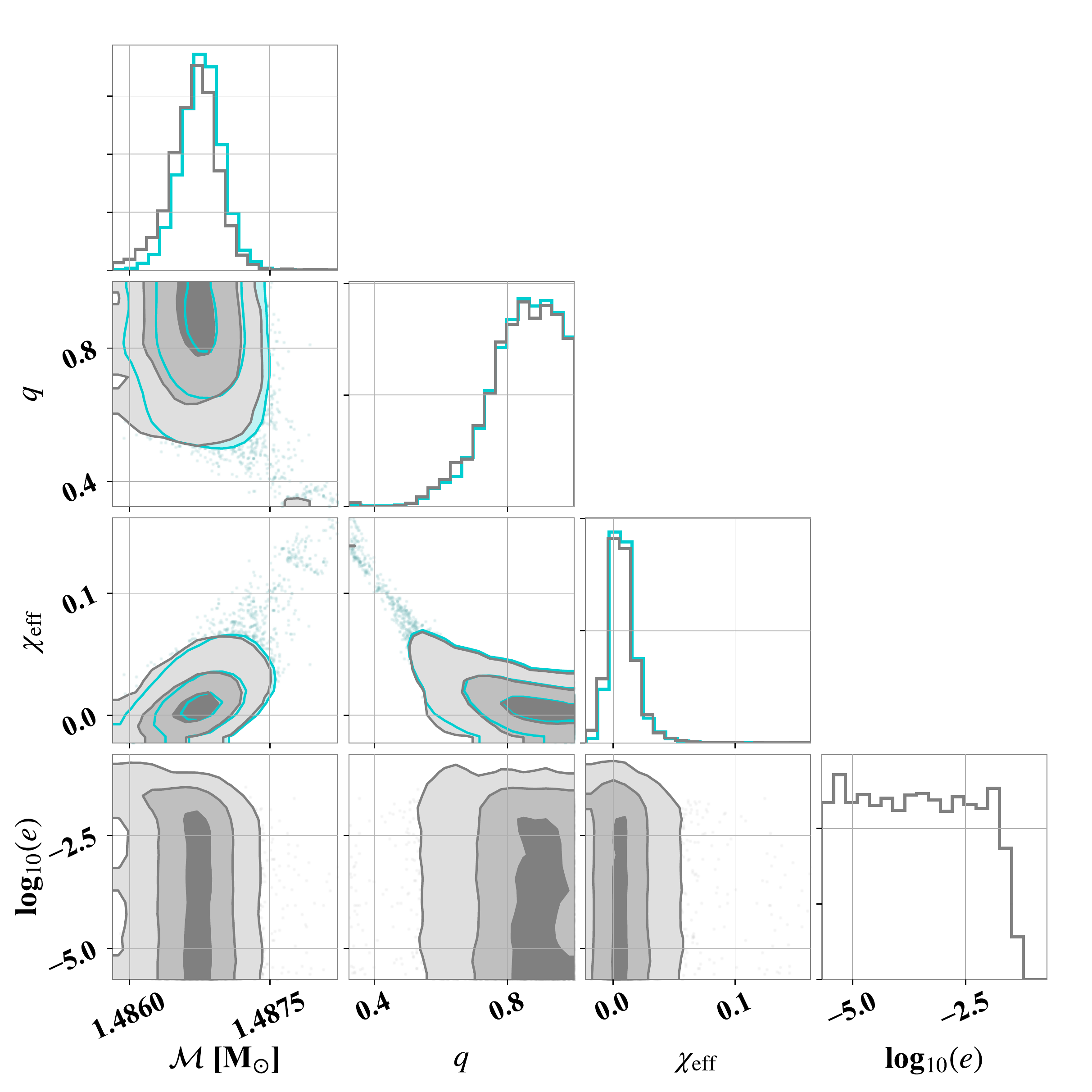}
    \caption{Recovered posterior probability distributions for GW190425 for intrinsic parameters: chirp mass $\mathcal{M}$, mass ratio $q$, effective spin $\chi_\mathrm{eff}$, and log eccentricity $\log_{10}(e)$. We plot the proposal posteriors in turquoise and the reweighted posteriors in gray.}
\end{figure}

 We present the posterior probability distributions obtained for a selection of intrinsic parameters for GW190425 in Fig.~\ref{fig:intrinsic}.

% \begin{figure*}
%     \label{fig:extrinsic_intrinsic}
%     \centering
%     \begin{subfigure}
%         \centering
%         \includegraphics[scale=0.35]{test_intrinsic_corner.pdf}
%     \end{subfigure}
%     ~ 
%     \begin{subfigure}
%         \centering
%         \includegraphics[scale=0.35]{test_extrinsic_corner.pdf}
%     \end{subfigure}
%     \caption{Recovered posterior probability distributions for GW190425. Intrinsic parameters chirp mass $\mathcal{M}$, mass ratio $q$, effective aligned spin $\chi_\mathrm{eff}$, and log eccentricity log$_{10}(e)$ are plotted on the left. Extrinsic parameters luminosity distance $d_\mathrm{L}$, binary inclination angle $\theta_\mathrm{jn}$, polarisation angle $\psi$, and orbital phase $\phi$ are plotted on the right. We plot the proposal posteriors in turquoise and the reweighted posteriors in gray.}
% \end{figure*}

\end{appendix}
\label{lastpage}
\end{document}